\documentclass[aps,nofootinbib,longbibliography,prd,twocolumn]{revtex4-1}
\usepackage[babel]{csquotes}
\usepackage{graphicx}
\usepackage{amsmath,amssymb}
\usepackage[colorlinks,citecolor=blue,linkcolor=blue,urlcolor=blue]{hyperref}
\usepackage{mathrsfs}
\usepackage{enumerate}
\def\be{\begin{equation}}
\def\ee{\end{equation}}

\begin{document}

\title{Natural discrete differential calculus in physics}

\author{Carlo Rovelli$^a$ and V\'aclav Zatloukal$^{a,b}$ \vspace{1mm}}

\affiliation{\small\mbox{$^a\ $CPT, Aix-Marseille Universit\'e, Universit\'e de Toulon, CNRS, F-13288 Marseille, France,\\ 
and}
}
\affiliation{$^b\ $FNSPE, Czech Technical University in Prague, 
B\v{r}ehov\'{a} 7, 115 19 Praha 1, Czech Republic
}
\date{\small\today}

\begin{abstract}
\noindent We sharpen a recent observation by Tim Maudlin: differential calculus is a natural language for physics only if additional structure, like the definition of a Hodge dual or a metric, is given; but the discrete version of this calculus provides this additional structure for free. 
\end{abstract}
\maketitle

\section{Introduction}
Field theory is the tool of choice of contemporary fundamental physics. Differential calculus is a natural language for field theory. Differential calculus is defined by a smooth manifold $\cal M$, the spaces of differential $p$-forms on this manifold, and the unique differential operator $d$.  This is a minimal mathematical structure, with a simple physical interpretation: $p$-forms represent local degrees of freedom and the manifold codes the adjacency relations between them.  

But differential calculus \emph{alone} is in general insufficient for physical theories. Additional structure is required. The minimal structure required is an operator $d^*$. This can be defined from a Hodge dual $*$, by
\be \label{dStar}
d^*=*\ d\ *,
\ee
or from a metric on the manifold, from which the Hodge dual can be defined.\footnote{It is conventional to include an additional sign in definition \eqref{dStar}, which depends on the manifold dimension, metric signature, and the rank of the differential form.}  
For instance, Maxwell theory can be written  in four dimensional language in terms of a 1-form $A$ (the potential), a 2-form $F$ (the field) and a 1-form $J$ (the source), satisfying
\be
dA=F,\ \   d^*F=J. 
\ee
The operator $d^*$ is called the codifferential; it  lowers the rank of the differential form. Similarly, in its three dimensional formulation, Maxwell theory can be given in terms of the time evolution of a 2-form $E$ (the electric field), a 1-form $B$ (the magnetic field), a 3-form $\rho$ (the charge), and a 2-form $j$ (the current) satisfying
\be
 d^*\!E=-\dot B, \ \   dE=\rho,\ \    d^*\!B=0,\ \    dB=j+\dot E,  \label{m3}
\ee
where the dot indicates time derivative.\footnote{In literature (for instance \cite{Frankel:2003fk}) one often encounters a dualized version of these equations, $$dE=-\dot B, \ \    d^*E=\rho,\ \   dB=0,\ \    d^*B=j+\dot E,$$ where $E$ is now a 1-form, $B$ a 2-form, $\rho$ a 0-form, and $j$ a 1-form.}  As another example, the Laplace(-deRham) operator, which is ubiquitous in physics, can be written as 
\begin{equation}
\Delta = (d+d^*)^2 = d d^* + d^* d.
\end{equation} 
In current fundamental physics, the interpretation of the metric, and the Hodge operator it defines, is provided  by general relativity: they are determined by the gravitational field, which is itself a dynamical variable.  This fact fogs the physical meaning of the mathematical structure of field theory, because the basic structure needed to define it is itself dynamical and ultimately quantum. 
 
On the other hand, however, quantum gravity indicates that spacetime is described by a smooth manifold \emph{only} within an approximation valid at scales larger than the Planck scale.  At small scale, we expect some form of discreteness to appear.  A number of current approaches to quantum gravity either assume (for instance \cite{Sorkin:1990bj}) or derive (for instance \cite{Rovelli1994a}) this discreteness.  

Tim Maudlin has recently pointed out that discreteness is sufficient to fill the missing ingredient for providing a complete mathematical language for elementary physics, based only on contiguity and local degrees of freedom \cite{Maudlin2018}. The ingredient that differential calculus misses is instead naturally present in the discrete version of differential calculus. Here we show how this is possible in a rather straightforward manner. 

\section{From continuous to discrete fields}

Examples of discrete versions of field theory are lattice Yang Mills theory \cite{Wilson1974}, which plays an important role for computations in the physics of the strong interactions, and the spin network formalism in loop quantum gravity \cite{Rovelli2015}. In the first, the lattice spacing is taken to zero to obtain physical results; in the second, there is no lattice spacing and discreteness reflects the actual discreteness of physical spacetime. 

In either case, the intuitive relation between the discrete and the continuum theory can be understood as follows.  Consider a triangulation $\Delta$ of an $n$-dimensional oriented spacetime manifold, namely a partition of the manifold in $n$-simplices. These are bounded by oriented $(n-1)$-simplices (tetrahedra, if $n=4$), in turn bounded by oriented $(n-2)$-simplices (triangles, if $n=4$), until the zero-simplices, which are points of the manifold.   A $p$-simplex is characterised by the subset formed by the $p+1$ vertices in its boundary. Its boundary is the union of $(p+1)$ oriented ($p-1$)-simplices.  (For instance, the boundary of a tetrahedron is formed by four oriented  triangles.)

Since $p$-forms can be integrated over a $p$-dimensional surface, a physical $p$-form field $\phi$ can be integrated over a $p$-simplex $\sigma_p$ and hence assigns a number
\begin{equation}
\phi(\sigma_p)=\int_{\sigma_p} \phi
\end{equation}
to each $p$-simplex. Namely it defines a map from the set of the $p$-simplices to $\mathbb{R}$.  This is a ``discrete field".  1-forms, for instance, are integrated on segments, 2-forms on triangles and so on; hence a discretized 1-form is an assignment of a number to each oriented segment, a discretized 2-form is an assignment of a number to each oriented triangle, and so on. 
 
A $p$-simplex $\sigma_p$ is characterised by the set $\sigma_p=(v_0,v_1,...,v_p)$ of its vertices.  The order of vertices defines orientation of the simplex --- swapping two vertices results in a minus sign. It is useful to introduce formal linear combinations of $p$-simplices. These are called $p$-chains, their set denoted by $C_p$. The boundary of a $p$-simplex $\sigma_p$ is the element of  $C_{p-1}$ defined by
\begin{equation}
\partial \sigma_p
= \sum_{j=0}^{p} (-1)^{j} (v_0,\ldots,v_{j-1},v_{j+1},\ldots,v_p) .
\end{equation}
One sees immediately that $\partial \partial = 0$ (the boundary of a boundary is empty).  

Discrete fields are then defined as linear functions on $p$-chains with values in $\mathbb{R}$, their set denoted by $C^p$. These are called cochains in the mathematical literature. We denote $\phi(\sigma_p)$ the value of a $p$-cochain $\phi$ on the single $p$-simplex $\sigma_p$.  The boundary operation $\partial: C_p \rightarrow C_{p-1}$, induces a corresponding operation on cochains:
\begin{equation}
d: C^p \rightarrow C^{p+1}
\quad,\quad
d\phi(\sigma_{p+1}) = \phi(\partial \sigma_{p+1}),
\end{equation}
and it is easy to see that this is the discrete version of the continuous $d$ operator.  It satisfies $dd = 0$. Intuitively,  relations between fields that vary little at the scale of the triangulation are satisfied or approximated by their discrete version.  For instance, the first of the equations in \eqref{m3} equates the integral of the charge density (3-form) on a tetrahedron (3-simplex) to the flux of the electric field (2-form) across its boundary (linear combination of triangles).

\section{Discrete field theory}

The abstract combinatorial structure formed by the set of the vertices and by the family of the subsets of vertices characterising each $p$-simplex, together with their boundary relations, defines an abstract combinatorial complex $\mathcal{K}$.  This complex $\mathcal{K}$, and its operator $d$ acting on its cochains can be considered by itself, with no reference to a continuous manifold \cite{Frankel:2003fk}.   It provides a sufficient structure for defining the discrete field theory.   

{ This structure has been studied in the literature in different contexts, see for instance  \cite{marsden,marsden2,ilpadrino,teixeira,goeck}.}

The main observation in this paper is that unlike for its continuous counterpart, this discrete structure comes naturally equipped \emph{also} with a $d^*$ operator, without need of choosing additional structure.  Hence discrete differential calculus, unlike its continuous counterpart, is sufficient for physics, with no need of additional structure.  The intuitive reason for this is that  discreteness implicitly defines a scale, absent in the continuous theory.  

To show this, notice first that unlike the continuous case, the discrete case carries a naturally defined inner product between $p$-cochains. The inner product of cochains $\phi, \psi \in C^p$ is defined as
\begin{equation} \label{InnerProd}
(\phi,\psi) = \sum_{\sigma_p} \phi(\sigma_p) \psi(\sigma_p) ,
\end{equation}
where the sum runs over all $p$-simplices in $\mathcal{K}$.  { (Notice that this product is different from the product defined for instance in \cite{marsden}, or by equation (25) in \cite{ilpadrino}, because it is determined by the discrete structure itself and does not depend on Riemannian, or geometrical, notions like a volume form.)}
This allows us to define the operator $d^*$ as the adjoint of $d$:
\begin{equation} \label{Adjoint}
(\phi,d\psi) = (d^*\phi,\psi)
\end{equation}
for any $\phi \in C^p$ and $\psi \in C^{p-1}$. This operator is the discrete version of the continuum $d^*$ operator.  The Laplace(-deRham) operator on cochains is given by
\begin{equation}
\Delta: C^p \rightarrow C^p
\quad,\quad
\Delta = (d+d^*)^2 = d d^* + d^* d.
\end{equation} 
The triple ($\mathcal{K}, d, d^*$) provides a discretization of the triple (${\cal M}, d, d^*)$. But while in the second $d^*$ is not  determined by ($\mathcal{M}, d$), in the first $d^*$ is determined by ($\mathcal{K}, d$).

\section{The $d^*$ operator}

To get clarity on the nature of the discrete operator $d^*$ (acting on cochains, namely discrete fields), we show that it is related to an operator $\partial^*$ (acting on chains, namely linear combinations of simplices), in the same manner $d$ is related to the boundary operator $\partial$. 

We define the operator $\partial^*$ acting on chains by
\begin{equation}
\partial^*:C_{p-1} \rightarrow C_p
\quad,\quad
d^*\phi(\sigma_{p-1})=\phi(\partial^*\sigma_{p-1}) .
\end{equation}
While the boundary operator $\partial$ decreases the dimension of a simplex it acts on, the operator $\partial^*$ increases it. 

Its explicit form can be worked out as follows. First, for any simplex $\sigma_p$, it is convenient to define a corresponding cochain $\bar{\sigma}_p$ with value 1 on $\sigma_{p}$ and zero elsewhere. With this we can write $\phi(\sigma_p) = (\phi,\bar{\sigma}_p) = (\bar{\sigma}_p,\phi)$.  Choosing $\phi = \bar{\sigma}_p$ and $\psi = \bar{\sigma}_{p-1}$ in Eq.~\eqref{Adjoint} gives a useful relation
\begin{equation} \label{ddStar}
d\bar{\sigma}_{p-1}(\sigma_p)
= d^*\bar{\sigma}_p(\sigma_{p-1}) .
\end{equation}
Using this, we have
\begin{eqnarray} \label{DualB}
\partial^*\sigma_{p-1}
&=& \sum_{\sigma_p} d^*\bar{\sigma}_p(\sigma_{p-1}) \sigma_p \nonumber\\
&=& \sum_{\sigma_p} d\bar{\sigma}_{p-1}(\sigma_p) \sigma_p \nonumber\\
&=& \sum_{\sigma_p} \bar{\sigma}_{p-1}(\partial\sigma_p) \sigma_p \nonumber\\
&=& \sum_{v} (v,\sigma_{p-1}) ,
\end{eqnarray}
where the last sum runs over vertices $v$ such that $(v,\sigma_{p-1})$ is a $p$-simplex of $\mathcal{K}$.  That is: the $\partial^*$ of a ($p-1$)-simplex is formed by all the $p$-simplices that have the ($p-1$)-simplex in their boundary. Below we will see that this has a nice geometrical and combinatorial interpretation in terms of the dual complex. {  The simple combinatorial expression for $d^*$ given in the last line of equation \eqref{DualB} is a key result of this note.  In fact, we could have taken it as a purely combinatorial definition of the operator $\partial^*$, based only on contiguity in the simplicial complex. }

In passing we note that the latter equations imply 
\begin{equation} \label{CoboundDual}
d \bar{\sigma}_p
= \overline{\partial^*\sigma_p} ,
\end{equation}
where the ``bar" operation has been extended from simplices to all chains by linearity.

\begin{figure}[t]
\label{fig:complex}
\includegraphics[scale=1]{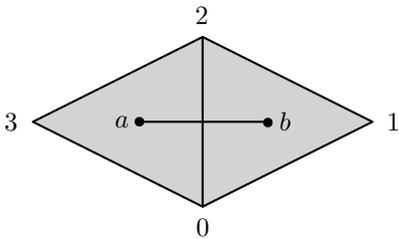}
\caption{\em \small Example of a complex $\mathcal{K}$. According to Eq.~\eqref{DualB}, we have $\partial^*(0,2) = (1,0,2)+(3,0,2)$. In the dual complex approach we identify the dual vertices $a=*(0,2,3)$ and $b=*(0,1,2)$, and the dual edge $*(0,2)=(b,a)$. Eq.~\eqref{BoundDualCompl} then gives $\partial^*(0,2) = *^{-1} \partial *(0,2) = (0,2,3)-(0,1,2)$, which coincides with the former result upon reordering of vertices.}
\end{figure}

With the chain equivalent $\partial^*$ of the discrete codifferential $d^*$ at our disposal, we easily find the chain equivalent of the Laplace operator by writing $\Delta \phi(\sigma_p) = \phi((\partial^* \partial + \partial \partial^*)\sigma_p)$.
For example, on $0$-cochains, i.e., functions defined on vertices, the Laplace operator reads
\begin{eqnarray}
\Delta \phi(\sigma_0)
&=& \phi(\partial \partial^*\sigma_0) \nonumber \\
&=& \sum_v \phi(\partial (v,\sigma_0)) \nonumber \\
&=& {\rm deg}(\sigma_0) \phi(\sigma_0) - \sum_v \phi(v) ,
\end{eqnarray}
where ${\rm deg}(\sigma_0)$ is the degree (the number of neighbours $v$) of the vertex $\sigma_0$.

\section{The dual complex}

The duality described above can be made explicit by introducing the dual complex.  To illustrate it, let us return for simplicity to the case of the triangulated oriented manifold.  If $\mathcal{K}$ is a triangulation of an orientable $n$-dimensional manifold $\mathcal{M}$, then there exists a \emph{dual} cellular complex $\mathcal{K}^*$, which offers a convenient dual interpretation of the operator $\partial^*$, analogous to the continuum definition in Eq.~\eqref{dStar}. 

The dual complex $\mathcal{K}^*$ can be defined as follows. (For complete details, see \cite{Munkers}.)   By definition, the vertices of $\mathcal{K}^*$ are the $n$-simplices of $\mathcal{K}$ and, in general, the $p$-cells of $\mathcal{K}^*$ are the $(n-p)$-simplices of $\mathcal{K}$.  Remarkably,  the $\partial^*$ operator of $\mathcal{K}$ turns out to be the boundary operator on $\mathcal{K}^*$.  See Figure 1 for an example. 

The dual map $*$ that sends a $p$-simplex in the corresponding $(n-p)$-cell and viceversa can be regarded as a discrete version of the continuum Hodge operator, and allows us to write
\begin{equation} \label{BoundDualCompl}
\partial\ * = *\ \partial^* ,
\end{equation} 
i.e., the operator $\partial^*$ is simply the operator dual to the boundary operator in the dual complex.
Observe the analogy with Eq.~\eqref{CoboundDual}, which relates the operator $\partial^*$ on chains, and the discrete differential operator $d$ on cochains via the ``bar" operation.

\section{Conclusions}

Continuous structures can be approximated by discrete structures with many elements (as in lattice QCD). Discrete structures with many elements can be approximated by continuous ones (as in the hydrodynamical approximation of the molecular dynamics of a liquid).  The question whether it is continuity or discreteness to be `more fundamental' has been posed repeatedly during  the history of physics, with oscillating answers. In his inaugural lecture at G\"ottingen in 1854, Bernhard Riemann famously observed that ``in a discrete manifold, the ground of its metric relations is given in the notion of it, while in a continuous manifold, this ground must come from outside."  \cite{br}\   On a similar vein, we have pointed out here that the minimal structure needed to define physical field theories is not provided by differential calculus alone, but is provided by the corresponding discrete structure: an abstract combinatorial complex.  

Abstract combinatorial complexes form the basic structure for the definition of the covariant, or spinfoam, formulation of the quantum dynamics of the gravitational field in loop quantum gravity \cite{Rovelli2015}.  The observation in this paper sheds some light on the role of such structure in this theory.

{ 

\vspace{1cm}
\centerline{---}
\vspace{1cm}

V.Z. acknowledges support from the ESIF, EU Operational Programme Research, Development and Education, and from International Mobility of Researchers in CTU (CZ.02.2.69/0.0/0.0/16\_027/0008465), Czech Technical University in Prague. C.R. acknowledges support from the SM Center on Space, Time and the Quantum.

}





\end{document}